\documentclass[a4paper]{jpconf}
\usepackage{graphicx}
\usepackage[utf8x]{inputenc}
\usepackage{wrapfig}
\usepackage{float}

\begin{document}
\vspace*{-1in}
\title{Stability of Metallic Hydrogen at Ambient Conditions}
\author{Graeme J Ackland}
\address{CSEC, SUPA, School of Physics and Astronomy, The University of Edinburgh, Edinburgh EH9 3JZ, United Kingdom}
\ead{gjackland@ed.ac.uk}

\begin{abstract}
The possibility of metallic hydrogen was first mooted by Wigner and
Huntington\cite{wigner1935possibility} in 1935. Here it is show that
the calculations from that paper are in remarkably good agreement with
modern density functional theory results. The possibility that
metallic hydrogen could be recovered to ambient pressure is often
attributed to papers by Brovman {\it et
  al}\cite{brovman1972properties,brovman2,brovman1}, although in fact
they only say it would be metastable with undetermined lifetime.
Density functional theory calculations presented here show that
reasonable candidate structures for metallic hydrogen are wildly
unstable at ambient conditions, and molecular dynamics calculations
show that the lifetime to which Brovman {\it et al} refer is
considerably less than a picosecond. It is concluded that the
prospects of using recovered metallic hydrogen as rocket fuel or for
electricity distribution may have been overstated.
\end{abstract}

\section{Introduction}

In 1926 J. D. Bernal proposed that all materials, when subjected to a
high enough pressure, will become metallic.  Several years later,
Wigner and Huntington investigated the properties of a putative
metallic modification of hydrogen.  Their idea was that hydrogen would
become atomic under pressure, and therefore behave like an alkali
metal.  The initial calculation by Wigner and Huntington (WH)\cite{wigner1935possibility} involved a
full potential soution to the Schroedinger equation within a radius
$r_s$, augmented by a free-electron contribution.  This calculation
gives a density of 0.8 at ambient pressure and an energy much higher
than molecular hydrogen, and so they concluded that the metallic phase
could exist only at high pressure.  A number of corrections for
exchange, correlation, Madelung, zero-point nuclear fluctuations
etc. in various approximations are considered, all of which increase the
instability at ambient pressure.

\begin{wrapfigure}{r}{80mm}
% \begin{figure}[H]
\vspace{-0.8cm}
\includegraphics[width=75mm]{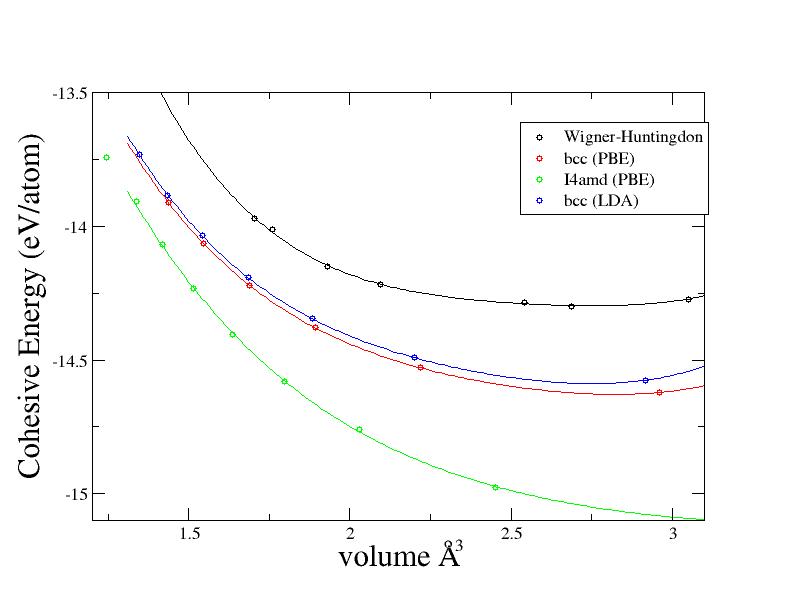}
\caption{Comparison of data transcribed from WH for the bcc phase, with CASTEP-PBE
  calculation for bcc and I4amd hydrogen and CASTEP-LDA for bcc
  hydrogen.}
\label{HW}
% \end{figure}
\vspace{-0.5cm}
\end{wrapfigure}

The physics included in these calculations is rather similar to that
still used in density functional calculations, although modern
computing power enables us to calculate the wavefunctions
self-consistently.  The comparison between WH's calculation and a
modern DFT calculation for bcc hydrogen is shown in Figure\ref{HW},
where one sees immediately that the agreement is rather good.  This
may be a surprise, since one commonly sees statements such as ``Wigner
and Huntington predicted a transition pressure of 25GPa''.  In fact
this 25GPa value is given as a lower bound, presented by WH to
demonstrate the impossibility of making metallic hydrogen with
contemporary equipment: WH also say ``if the compressibility at
ordinary pressures would hold throughout, the molecular form would be
stable for all volumes'' (i.e. the transition pressure is {\it infinite}).
WH presciently mention layered structures, but their considered transition 
goes directly from a molecular
insulator to atomic free-electron material, and in the most recent
report of metallic hydrogen, Dias and Silvera
\cite{SilveraMetalScience} show a reflectivity corresponding to such a
free electron metal.  However most DFT calculations suggest that
metallisation may first occur via band-gap closure in a molecular
solid, with the atomic solid appearing at higher pressures.

The exciting properties claimed for metallic hydrogen, such as room
temperature superconductivity, are expected in the atomic phase, so it
is the recovery of this phase which we consider.  The idea of
recoverability of atomic hydrogen is typically associated with the
1974 work of Brovman {\it et al}\cite{brovman1}.  They apply a model
similar to WH, with a perturbative approach and
simplifying assumptions, e.g. {\it ``the correlation energy has
  practically no influence on the determination of the optimal
  structure''}.  They test a number of possible crystal structures and
find that the Wigner-Huntington-type monoatomic structures tend to be
unstable with respect to unit cell doubling (a precursor to molecule
formation).  Like WH, they find that the metallic
phases have an energy minimum corresponding to zero pressure and state
that there should be a {\it ''phase locally stable in all the
  macroscopic parameters''}, which is stable relative to atomisation
but has higher energy than the molecular phases.  Unfortunately, many
readers have failed to notice that the caveat ``macroscopic
parameters'' involves only a limited range of possible instabilities.
They consider only affine deformations (i.e. elastic instabilities), and
not phonons.  This has led to a
misapprehension that Brovman's work predicts a recoverable metallic
state, ignoring Brovman's subsequent statement that the lifetime of
this state {\it ``remains open''}.

Perhaps the most curious aspect of this is why Brovman's work is still
regarded as plausible when lattice dynamics calculations which
probe both microscopic and macroscopic deformations have been routine
for many years.  Self-consistent methods of calculating electronic
structure, in particular DFT, are able to treat both metallic and
covalent bonding within the same microscopic theory.  Chemically, one
would expect the instability of atomic hydrogen to be towards
formation of molecules, however the tests of stability against affine
deformations carried out by Brovman do not allow such pairing in dense
hydrogen.

The question of finding the most stable arrangement of atoms has been
tackled in the seminal paper by Pickard and
Needs\cite{pickard2007structure}, and subsequently by a number of
other authors\cite{geng2012high,monserrat2016hexagonal}.  From this work, the most likely 
candidate structure
for atomic metallic hydrogen has emerged as the four-fold coordinated
$I4amd$.

\section{Calculation Details}
Energies and phonons were calculated for WH's $bcc$ phase and the
$I4amd$ candidate phase for various pressures. The
CASTEP\cite{segall2002first} code was used with the PBE functional and
other settings as used
previously\cite{magdau2013identification,ackland2015appraisal,magdau2017infrared}. Phonon
calculations were done using 0.01bohr finite
displacements\cite{SUPERCELL_PHONON,ackland1997practical} with around
3000 k-points per primitive cell,  which still introduced a sampling error
around 50$cm{^-1}$.  Energy calculations exclude the zero-point energy
and pressure.

The $I4amd$ symmetry can refer to two very different structures
depending on the c/a ratio.  Exemplars for this are $\beta-$Sn and
Cs-IV which have c/a ratio in the conventional 4-atom cell around 0.6
and 2.2 respectively.  For hydrogen the Cs-IV type is generally
regarded as most stable\cite{geng2012high} across a range of exchange
correlation functionals\cite{clay2014benchmarking,morales2013towards}.

There is insufficient detail in the Wigner and Huntington paper to
exactly reproduce their calculations, so the data was transcribed
directly from Figure 3 in that paper, with units converted 
to eV and $\AA^3$.

The atomic metallic $I4amd$ is a stable representative of the
free-electron WH phase\cite{geng2012high}, we examined the phonons as
a function of pressure.  McMahon {\it et al}\cite{mcmahon2011ground} have
presented the pressure dependence of the $\Gamma$ point phonons, which
were postulated as a match for Phase V Raman
data\cite{dalladay2016evidence}, however, the validity of this is
unclear, since five non-zero $\Gamma$ point frequencies are presented
from a structure with only two atoms in the primitive cell.  Probably
they ignored the body centring, an error replicated in Fig 2b.

Although $I4amd$ is stable with respect to
phonon distortions above 250GPa, it develops a fully unstable phonon
branch at ambient conditions.  Such an instability across the entire
Brillouin zone means the structure is unstable to both localised and
phonon distortions. Allowing these distortions in molecular dynamics
started at 10K takes the structure to an insulating molecular
structure within tens of femtoseconds. We do find that the acoustic
branches are all positive, consistent with the metastability checks
of elastic moduli carried out by Brovman et al.

%\begin{wrapfigure}{R}{90mm}
 \begin{figure}[H]
\includegraphics[width=50mm]{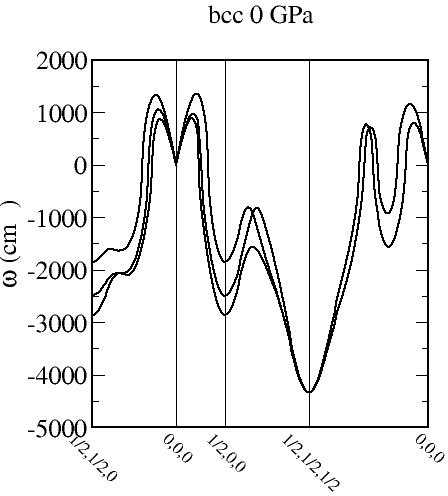}
\includegraphics[width=50mm]{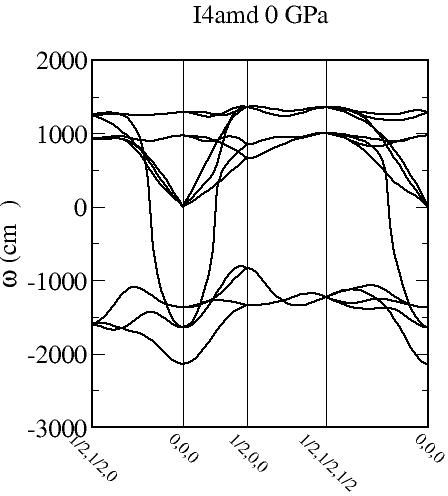}
\includegraphics[width=55mm]{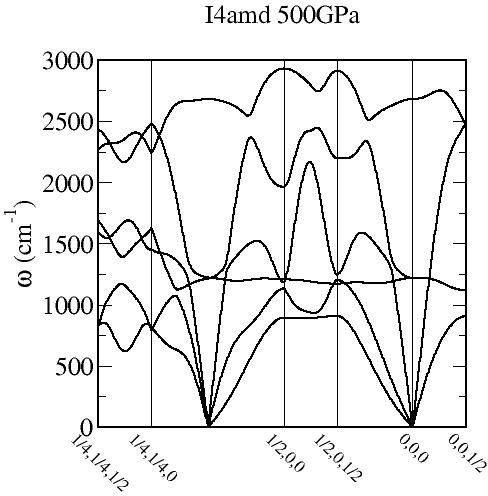}
\caption{Phonon dispersion calculated for the WH bcc phase at 0 GPa
  and for $I4amd$ metallic hydrogen at 0 and 500GPa.  The 0GPa
  calculation takes the 4 atom tetragonal $I4amd$ unit cell, while the
  500GPa used the primitive cell with the two Raman active modes Eg at 
 1218$\pm50cm^{-1}$. and B2g at  2679$\pm50cm^{-1}$. 
 Imaginary frequencies are shown as negative}
\label{I4amd-PBE}
 \end{figure}
%\end{wrapfigure}

\section{Results and discussion}

\begin{wrapfigure}{r}{70mm}
% \begin{figure}[H]
\includegraphics[width=70mm]{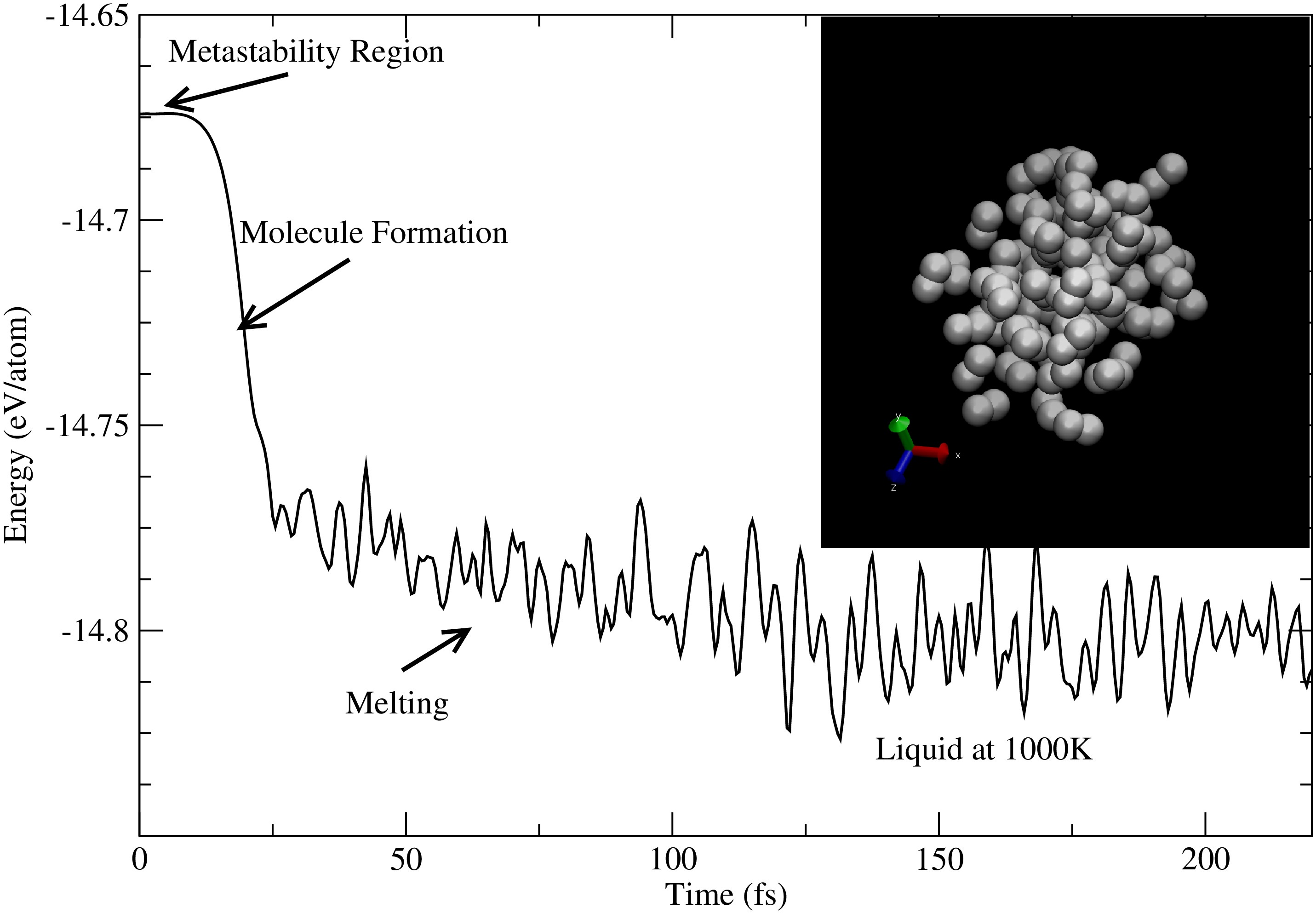}
\caption{Molecular dynamics energy as a function of time for metallic
  hydrogen, initialized as $I4amd$ at 10K and 0GPa. Inset shows a
  snapshot showing molecular liquid after 70fs.}
\label{bang}
% \end{figure}
\end{wrapfigure}

The graphs of energy and volume show remarkable agreement of the
WH calculations and modern DFT.  
WH calculate that the minimum energy
for bcc at $-1.05Ry$, $r_s=1.63a.u.$ compared to the modern values of
$-1.075Ry$ and $r_s=1.6834a.u.$, a level of accuracy not dissimilar to the
scatter of modern density functionals.
The $I4amd$ structure is significantly more stable than
bcc at all volumes.  In fact, WH's error in
choosing  bcc rather than $I4amd$ is larger than that due to their
approximations to electronic structure calculation.

To test how long metallic hydrogen remains metastable, a 108 atom
supercell of the $I4amd$ structure was relaxed to ambient pressure while
enforcing symmetry, then {\it ab initio} molecular dynamics was
started at 10K in the NVE ensemble. Figure \ref{bang} shows the
evolution of the energy through four distinct phases, 15fs of I4amd
metastability, followed by a further 15fs of molecule formation.
 We implemented a molecule-finding algorithm which associates each atom
with its nearest neighbour, and thereby detects whether a unique
definition of diatomic molecules exists.  After 38fs such a configuration was
found.  After that some making and breaking of molecules continued up
to 100fs, when the material equilibrated as a liquid and temperature
had risen to 1000K with a pressure increase of 40GPa.  After this the
structure was a liquid molecular insulator, with occasional rebonding of the
molecules occurring.

\section{Conclusion}
The predictions of density functional theory for bcc metallic hydrogen
are compared with the analytic results obtained by Wigner and
Huntington in 1935.  The agreement between the two calculations is
within a few percent for energy and Wigner Seitz radius, the
compressibility in the two calculations is also similar.  This
remarkable result seems not to have been properly appreciated, with
modern authors still repeating the claim that WH ``predicted a
transformation at 25GPa''.  The difficulty for WH in calculating and
accurate transition pressure is that their theory was unable to treat
molecular phases on the same footing as the metallic ones.  In
retrospect, they would have been wise to avoid making any claims about
stability.  

The elastic stability of the best known candidate structure for the
atomic phase, $I4amd$\cite{geng2012high}, has been investigated.  It
passes the metastability tests considered by Brovman, having positive
acoustic phonons and harmonic elastic moduli at all pressures.
However, we find massive optical phonon instabilities leading to
molecularization at ambient pressure.

Molecular dynamics provides the answer to Brovman's issue about the {\it
  ``remains open''} lifetime of the recovered state: in the most
advantageous case where the depressurization is instantaneous, $I4amd$
metallic hydrogen at ambient pressure has a lifetime of order 40fs.
The power generated during molecularisation is calculated to be an
impressive 10$^{17}$ W/mol, but the fuel loading and short release
timescale would prove challenging for practical rocketry.

In summary, the predictions of two classic papers on metallic hydrogen
at ambient pressure have been reexamined.  The energy,
volume and compressibility calculated by WH for bcc hydrogen is in
remarkably good agreement with DFT calculations.  Their prediction
that the transition pressure lies somewhere between 25GPa and infinite
pressure is also vindicated.  The work of Brovman fares less well:
although we find that their calculations showing metastability of
metallic hydrogen against affine deformation are valid, the stricter
criterion of phonon stability is not met.  As a consequence, the
expected lifetime of putative metallic hydrogen at ambient condition can be
measured in femtoseconds.

I acknowledge support from the ERC Hecate and the EPSRC UKCP projects.

\vspace{5mm}

\bibliographystyle{iopart-num}
\bibliography{Refs}

\end{document}